\documentclass[twocolumn,showpacs,preprintnumbers,amsmath,amssymb]{revtex4}

\usepackage{graphicx}
\usepackage{dcolumn}
\usepackage{bm}
\usepackage{color}

\usepackage{graphicx}
\usepackage{dcolumn}
\usepackage{bm}

\begin{document}

\preprint{}
\input{epsf.tex}

\epsfverbosetrue

\title{Dark Bogolon-Excitons in a Linear Atomic Super-Lattice}
\author{Hashem Zoubi}

\affiliation{Max-Planck Institute for the Physics of Complex Systems, Noethnitzer Str. 38, 01187 Dresden, Germany}

\date{14 November 2014}

\begin{abstract}
Dark and bright excitons are shown to appear naturally in a linear atomic
super-lattice with two atoms per unit cell. In bringing the super-lattice into
a strong coupling regime with a one-dimensional nanophotonic waveguide, bright excitons and
photons are coherently mixed to form polaritons. Treating excitons as bosons implies a
mechanism that forbids two excitations from being at the same atomic state,
which is included here through a bosonization procedure with kinematic
interactions. Interestingly these interactions couple dark and bright
excitons, and which we exploit as a new tool for exciting dark states in a controllable way. We
suggest a pump-probe experiment where two polaritons scatter into two dark
excitons that found to be
correlated and are represented as dark bogolon-excitons. The results can be
adapted for any super-lattice of active materials, e.g., of organic molecules.
\end{abstract}


\maketitle

\section{Introduction}

Collective electronic excitations (excitons) have been a matter of research since their
introduction by Frenkel in 1931 \cite{Frenkel}. Frenkel excitons are of
importance for fundamental physics, as they explain various electrical and
optical properties in molecular crystals, e.g. in organic and noble atom crystals
\cite{Agranovich,Davydov}. These type of excitations have been
extensively investigated and found applications for
optoelectronic devices \cite{Pope,Zoubi2005a}. The mechanism behind their appearance is through the
delocalization of electronic excitations among the crystal atoms or molecules
via electrostatic interactions, mainly resonant dipole-dipole interactions,
and in exploiting the lattice symmetry. Such coherent states can appear
spontaneously among electronic states of a huge number of atoms without the aid
of external fields. Just to mention that coherence among electronic states inside
the same atom implies external laser fields, as for electromagnetic induced
transparency and population trapping \cite{Fleischhauer}.

The progress in the fabrication of optical lattices provides artificial
crystals with controllable parameters \cite{Jaksch,Greiner}, which are adopted
to simulate a wide range of complex condensed matter models and present a deep understanding of puzzling
solid state effects \cite{Lewenstein}. Of importance here is ultracold atoms in optical
lattices at the Mott insulator phase with one atom per site, which have been proposed for
simulating Frenkel excitons and unveiled physical phenomena that
are masked by thermal excitations and impurities in solid crystals \cite{Zoubi2007,Zoubi2013}.

On the other hand optical lattices have been introduced as active material for
cavity QED \cite{Zoubi2007}. It was shown that electronic excitations and cavity photons in the
strong coupling regime are naturally mixed to form polaritons as the system collective
excitations \cite{Zoubi2013}. Several configurations of optical lattices within a cavity have
been suggested with diverse dimensionality and geometry \cite{Zoubi2008,Zoubi2009}. Of special
interest is that of a one dimensional optical
lattice formed by a nanophotonic waveguide \cite{NayakA,NayakB}, which was realized for cesium atoms
in using tapered nanofibers \cite{Vetsch,Goban}. The optical lattice is
localized outside and parallel to the nanofiber by evanescent fields
of counter propagating colored fiber beams. The atomic lattice
can be directly interrogated by a resonant beam that sent through the fiber and
observed via linear optical spectra \cite{Zoubi2010}.

In the present paper we address an optical super-lattice with two sites per unit
cell and that includes one atom per site, where the distance between the two on
cell atoms is taken to be smaller than the distance
between the centers of two neighbor unit cells. This configuration gives rise
to symmetric and antisymmetric entangled states, in which an electronic
excitation can be delocalized among the two atoms inside the same unit cell. The
symmetric states are bright and superradiant with damping rate of twice that
of a single excited atoms. While the antisymmetric states are dark with zero
damping rate and decouple to the light \cite{Ficek}. It is a big challenge to excite dark states in an efficient and controllable way. In localizing the super-lattice parallel to
a nanophotonic waveguide at a fixed distance, only the symmetric excitons coherently couple to the
fiber photons in order to form one dimensional polaritons.

Excitons in an extended system behave as bosons, but then they acquire a mechanism
that forbids double excitations at the same atom \cite{Toshich,Zoubi2005b}. In previous work we presented
a bosonization procedure that transform two-level atoms into bosons with
additional kinematic interactions that forbid two excitations from being localized on
the same atom, and which found to be a good approximation at low
intensity of excitations \cite{Zoubi2014}. Here we apply such a procedure to optical
super-lattices, and we interestingly show that kinematic interactions can
couple symmetric and antisymmetric states. We exploit this mechanism as a
controllable mean for
exciting dark excitons. In order to achieve our goal we proposed a pump-probe
experiment, where two pumped polaritons with momentum $k$ can be scattering into two dark
excitons of momenta $k+p$ and $k-p$. This parametric amplification of dark
states is induced by a probe field
at either $k+p$ or $k-p$
momenta, and subjected to
conservation of energy. The scattering of two polaritons into a state of two excited atoms in
the same unit cell can be excluded due to conservation of energy. In the frame of
Bogoliubov mean field theory the dark exciton pairs are strongly
correlated and can be represented by bogolon excitations that
termed here bogolon-excitons \cite{Fetter}.

The paper is organized as follows. In section 2 we introduce dark and bright
excitons into an optical super-lattice. An atomic super-lattice strongly coupled to a
nanophotonic waveguide is presented in section 3 by using the language of
polaritons. Kinematic interactions are derived in section 4 with emphasize on
the symmetric-antisymmetric mixing. Pump-probe experiments appear in section 5
where a process for exciting dark states is proposed. Section 6 includes
Bogoliubov mean field theory with dark bogolon-excitons. A summary is given
in section 7.

\section{Collective Excitations in Linear Super-Lattices}

We consider a system of one-dimensional optical lattice with a basis of two
sites (super-lattice). Namely, we have a lattice with unit cell of two trapping
minima, as
seen in figure (1). The lattice constant is $a$, which is the distance between the center of two neighbor unit cells, and the
distance between the two sites within the same unit cell is $R$. The optical
super-lattice can be realized in using counter propagating laser beams to get a
standing wave super-lattice \cite{NayakA,NayakB}. Ultracold or cold atoms are loaded into such a system,
where the atoms experience optical super-lattice potential. We treat the case
of one atom per site with two atoms per unit cell. Here we
concentrate in a system of nanophotonic waveguide, as beside the off-resonance
light used to produce the super-lattice it provides a strong resonant coupling of light to the
atomic super-lattice. The recent set-up of tapered nanofibers have the
potential for realizing such a system \cite{Vetsch,Goban}.

\begin{figure}[h!]
\centerline{\epsfxsize=7cm \epsfbox{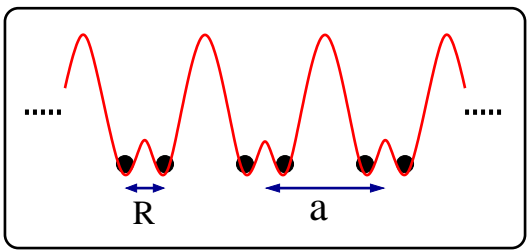}}
\caption{Optical super-lattice with two sites per unit cell and one atom per
  site. The on-site atoms are separated by $R$, and neighbor unit cells by $a$.}
\end{figure}

The atoms are taken to be of two-level systems with atomic transition energy
$E_A$. In using the second quantization language, the excitation Hamiltonian
reads
\begin{equation}
H_{ex}=E_A\sum_{n,\alpha}S_{n}^{\alpha\dagger}S_{n}^{\alpha}+\sum_{nm,\alpha\beta}J_{nm}^{\alpha\beta}\
S_{n}^{\alpha\dagger}S_{m}^{\beta},
\end{equation}
where $S_{n}^{\alpha\dagger}$ and $S_{n}^{\alpha}$ are the creation and
annihilation operators of an electronic excitation at atom $\alpha$ in site
$n$, respectively. Here $(\alpha,\beta=1,2)$ stand for the two atoms at the same unit
cell, and $(n,m=0,\pm 1,\cdots,\pm M)$ for the summation over the super-lattice
sites, with $N=2M+1$ the number of unit cells.

The Hamiltonian includes resonant dipole-dipole interactions between the atoms that
allow for energy transfer, in which an excited atom decays to
the ground state and other atom is excited. The interaction parameter among
$\alpha$ and $\beta$ atoms at sites $n$ and $m$ is $J_{nm}^{\alpha\beta}$,
which is
\begin{equation}
J_{nm}^{\alpha\beta}=\frac{\mu^2}{4\pi\epsilon_0(R_{nm}^{\alpha\beta})^3}\left(1-3\cos^2\theta\right),
\end{equation}
where $\mu$ is the transition dipole that has an angle $\theta$ with the super-lattice axis. The interatomic distance is
$R_{nm}^{\alpha\beta}=\left|R_{n}^{\alpha}-R_{m}^{\beta}\right|$, where
$R_{n}^{\alpha}$ is the position of atom $(n,\alpha)$ in the super-lattice.

We write the Hamiltonian as a sum of two terms, the first includes
interactions among atoms in the same unit cell, and the second
among different unit cells. We can
write $H_{ex}=\sum_{n}H_n+\sum_{nm}^{\prime}H_{nm}$, where
\begin{equation}
H_n=E_A\sum_{\alpha}S_{n}^{\alpha\dagger}S_{n}^{\alpha}+\sum_{\alpha\beta}^{\prime}J_{n}^{\alpha\beta}\
S_{n}^{\alpha\dagger}S_{n}^{\beta},
\end{equation}
and
\begin{equation}
H_{nm}=\sum_{\alpha\beta}J_{nm}^{\alpha\beta}\ S_{n}^{\alpha\dagger}S_{m}^{\beta}.
\end{equation}
The primes indicate that in the summations we have $(n\neq m)$ and $(\alpha\neq\beta)$.

The $H_n$ Hamiltonian can be diagonalized in using the transformation
\begin{equation}
S_{n}^s=\frac{S_{n}^1+S_{n}^2}{\sqrt{2}},\ S_{n}^a=\frac{S_{n}^1-S_{n}^2}{\sqrt{2}},
\end{equation}
which yields
\begin{equation}
H_n=E_s\ S_{n}^{s\dagger}S_{n}^s+E_a\ S_{n}^{a\dagger}S_{n}^a,
\end{equation}
where $E_s=E_A+J_0$ and $E_a=E_A-J_0$, with
\begin{equation}
J_0=\frac{\mu^2}{4\pi\epsilon_0R^{3}}\left(1-3\cos^2\theta\right).
\end{equation}
Note that due to the dependence of $J_0$ on $\theta$, the polarization direction can provide
a control parameter for the symmetric-antisymmetric splitting energy.

In using the above transformation into symmetric and antisymmetric
operators in $H_{nm}$, we get explicitly
\begin{eqnarray}
H_{nm}&=&\frac{1}{2}\left\{\left(J_{nm}^{11}+J_{nm}^{22}+J_{nm}^{12}+J_{nm}^{21}\right)\
  S_{n}^{s\dagger}S_{m}^s\right. \nonumber \\
&+&\left.\left(J_{nm}^{11}+J_{nm}^{22}-J_{nm}^{12}-J_{nm}^{21}\right)\
  S_{n}^{a\dagger}S_{m}^a\right. \nonumber \\
&+&\left.\left(J_{nm}^{11}-J_{nm}^{22}-J_{nm}^{12}+J_{nm}^{21}\right)\
  S_{n}^{s\dagger}S_{m}^a\right. \nonumber \\
&+&\left.\left(J_{nm}^{11}-J_{nm}^{22}+J_{nm}^{12}-J_{nm}^{21}\right)\
  S_{n}^{a\dagger}S_{m}^s\right\}.
\end{eqnarray}
Let us assume interactions only among nearest neighbor unit cells, which is a
good assumption in optical lattices. Between the
$n$
and $n+1$ sites we have
\begin{eqnarray}
J_{n(n+1)}^{11}&=&J_{n(n+1)}^{22}=\frac{\mu^2}{4\pi\epsilon_0a^{3}}\left(1-3\cos^2\theta\right), \nonumber \\
J_{n(n+1)}^{12}&=&\frac{\mu^2}{4\pi\epsilon_0(a+R)^{3}}\left(1-3\cos^2\theta\right), \nonumber \\
J_{n(n+1)}^{21}&=&\frac{\mu^2}{4\pi\epsilon_0(a-R)^{3}}\left(1-3\cos^2\theta\right).
\end{eqnarray}
We assume now the limit of $a\gg R$, then $H_{nm}\approx 2J\
S_{n}^{s\dagger}S_{m}^s$, where
\begin{equation}
J=\frac{\mu^2}{4\pi\epsilon_0a^{3}}\left(1-3\cos^2\theta\right).
\end{equation}
The antisymmetric states are on-site localized, and the mixing terms between the
symmetric and antisymmetric states vanish. Only the
symmetric states can be delocalized among the supper-lattice sites.

In the limit of $R\ll a$, the Hamiltonian reads
\begin{equation}
H_{ex}=\sum_{n}\left\{E_s\ S_{n}^{s\dagger}S_{n}^s+E_a\ S_{n}^{a\dagger}S_{n}^a\right\}+\sum_{nm}^{\prime}2J\ S_{n}^{s\dagger}S_{m}^s.
\end{equation}
At this point we aim to diagonalize the Hamiltonian relative to the site
indices, in applying the transformation
\begin{equation}
S_{n}^{\nu}=\frac{1}{\sqrt{N}}\sum_ke^{-ikz_n}S_{k}^{\nu},\ (\nu=s,a),
\end{equation}
where the position of the unit cell center is $z_n=an$, and the wavenumber $k$ takes the values $k=\frac{2\pi}{Na}p$ with $p=0,\pm
1,\cdots,\pm M$. Here, an electronic excitation is delocalized in the lattice where
it has the same probability to appear at every lattice site. We get
\begin{equation}
H_{ex}=\sum_{k}E_s(k)\ S_{k}^{s\dagger}S_{k}^s+E_a\sum_{k} S_{k}^{a\dagger}S_{k}^a,
\end{equation}
which contains symmetric and antisymmetric excitons. The antisymmetric states
are dispersion-less with the energy $E_a=E_A-J_0$. The symmetric states have
the dispersion relation $E_s(k)=E_s+J(k)$, where $J(k)=2J\sum_ze^{ikz}$. For
nearest neighbor interactions we obtain $J(k)=4J\cos(ka)$, and the symmetric
dispersion now reads $E_s(k)=E_A+J_0+4J\cos(ka)$. In place of the discrete
energy levels we get an energy band. Excitons are collective electronic excitations that behave as
quasi-particles \cite{Agranovich,Davydov}. The combination of electrostatic interactions and lattice
symmetry gives rise to energy waves that propagate with wave numbers that are good
quantum numbers.

\section{Super-Lattice Excitations Coupled to Photons}

The optical supper-lattice is localized parallel to the nano-photonic waveguide
at a distance $b$, as in figure (2). We consider one-dimensional propagating
photons that are represented by the Hamiltonian
\begin{equation}
H_{ph}=\sum_q\ E_{ph}(q)\ a_q^{\dagger}a_q,
\end{equation}
where $a_q^{\dagger}$ and $a_q$ are the creation and annihilation operators of
a photon of mode $q$, respectively. The wave number $q$ takes the values $q=2\pi l/L$, where
$l=0,\pm1,\pm2,\cdots,\pm\infty$, and $L$ is the waveguide length. The photon dispersion can be given by
\begin{equation}
E_{ph}(q)=\frac{\hbar c}{\sqrt{\epsilon}}\sqrt{q_0^2+q^2},
\end{equation}
where $\epsilon$ is an effective dielectric constant, and $q_0$ is the
confinement wave number.

\begin{figure}[h!]
\centerline{\epsfxsize=8cm \epsfbox{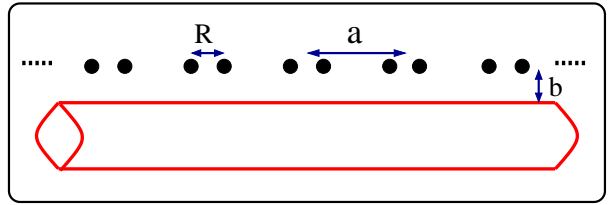}}
\caption{The optical super-lattice is localized parallel to a nano-photonic
  wave guide at distance $b$.}
\end{figure}

The excitation-photon coupling in the electric dipole approximation is written as $H_I=-\hat{\mbox{\boldmath$\mu$}}\cdot\hat{E}$, where the excitation
transition dipole operator is given by
$\hat{\mbox{\boldmath$\mu$}}=\mbox{\boldmath$\mu$}\sum_{n,\alpha}\left(S_{n}^{\alpha}+S_{n}^{\alpha\dagger}\right)$, and $\mbox{\boldmath$\mu$}$ is the transition dipole.

The photon electric field operator is given by
\begin{equation}
\hat{E}(r,z)=i\sum_q\sqrt{\frac{E_{ph}(q)}{2\epsilon_0V}}{\bf e}u(r)\left(a_qe^{-iqz}-a_q^{\dagger}e^{iqz}\right),
\end{equation}
where $V$ is the photon effective volume, ${\bf e}$ is the photon unit
vector polarization, and $u(r)$ is the photon mode function.

The interaction Hamiltonian in the rotating wave approximation, and in the Schr$\ddot{o}$dinger picture, reads
\begin{equation}
H_{I}=\sum_{qn,\alpha}\left(f_{qn}^{\alpha}\ a_qS_{n}^{\alpha\dagger}+f_{qn}^{\alpha\ast}\ S_{n}^{\alpha}a_q^{\dagger}\right),
\end{equation}
where
\begin{equation}
f_{qn}^{\alpha}=-i\sqrt{\frac{E_{ph}(q)}{2\epsilon_0V}}u(b)(\mbox{\boldmath$\mu$}\cdot{\bf
  e})e^{-iqz_n^{\alpha}}.
\end{equation}
The electric field is evaluated at the atom positions with $u(b)$ the mode
function at the lattice position. Here $z_n^{\alpha}$ is the position of atom $\alpha$ at site
$n$, where we write $z_n^{1}=z_n-\frac{R}{2}$ and $z_n^{2}=z_n+\frac{R}{2}$. In terms of symmetric and antisymmetric operators, we get
\begin{equation}
H_{I}=\sum_{qn,\nu}\left(f_{qn}^{\nu}\ a_qS_{n}^{\nu\dagger}+f_{qn}^{\nu\ast}\ S_{n}^{\nu}a_q^{\dagger}\right),
\end{equation}
with the coupling parameters
\begin{eqnarray}
f_{qn}^{s}&=&-i\sqrt{\frac{E_{ph}(q)}{\epsilon_0V}}u(b)(\mbox{\boldmath$\mu$}\cdot{\bf
  e})\left\{\frac{e^{-iqz_n^{1}}+e^{-iqz_n^{2}}}{2}\right\}, \nonumber \\
f_{qn}^{a}&=&-i\sqrt{\frac{E_{ph}(q)}{\epsilon_0V}}u(b)(\mbox{\boldmath$\mu$}\cdot{\bf
  e})\left\{\frac{e^{-iqz_n^{1}}-e^{-iqz_n^{2}}}{2}\right\}.
\end{eqnarray}
In momentum space we have
\begin{equation}
H_{I}=\sum_{k,\nu}\left(f_{k}^{\nu}\ a_kS_{k}^{\nu\dagger}+f_{k}^{\nu\ast}\ S_{k}^{\nu}a_k^{\dagger}\right),
\end{equation}
where
\begin{eqnarray}
f_{k}^{s}&=&-i\sqrt{\frac{E_{ph}(k)N}{\epsilon_0V}}u(b)(\mbox{\boldmath$\mu$}\cdot{\bf
  e})\cos(kR/2), \nonumber \\
f_{k}^{a}&=&\sqrt{\frac{E_{ph}(k)N}{\epsilon_0V}}u(b)(\mbox{\boldmath$\mu$}\cdot{\bf
  e})\sin(kR/2).
\end{eqnarray}
The interatomic distance $R$ can provide a significant control parameter for the
interaction of the photons to the symmetric and antisymmetric excitons. This mechanism can be
used to exchange the states from dark into bright and vice versa.

The total Hamiltonian now reads
\begin{eqnarray}
H&=&\sum_k\left\{E_{ph}(k)\ a_k^{\dagger}a_k+E_{s}(k)\
  S_{k}^{s\dagger}S_{k}^s+E_{a}\
  S_{k}^{a\dagger}S_{k}^a\right. \nonumber \\
&+&\left.\sum_{\nu}\left(f_{k}^{\nu}\ a_kS_{k}^{\nu\dagger}+f_{k}^{\nu\ast}\ S_{k}^{\nu}a_k^{\dagger}\right)\right\}.
\end{eqnarray}
Due to translational symmetry the Hamiltonian is separated in momentum
space, and the interaction is between excitons and photons with the same wave number.

In the limit of small $k$, that is $kR\ll 1$, we have
\begin{equation}
f_{k}^{s}\approx -i\sqrt{\frac{E_{ph}(k)}{\epsilon_0\bar{S}a}}u(b)\mu,\ f_{k}^{a}\approx \sqrt{\frac{E_{ph}(k)}{\epsilon_0\bar{S}a}}u(b)\mu\frac{kR}{2},
\end{equation}
where $\mu=(\mbox{\boldmath$\mu$}\cdot{\bf
  e})$. We used $V=\bar{S}Na$ and $L=Na$, with $\bar{S}$ the photon effective cross
section. The coupling of symmetric excitons to the photons is much larger than the
antisymmetric ones.

In neglecting the antisymmetric weak coupling part, the
Hamiltonian is written as
\begin{eqnarray}
H&=&\sum_k\left\{E_{ph}(k)\ a_k^{\dagger}a_k+E_{s}(k)\
  S_{k}^{s\dagger}S_{k}^s+E_{a}\
  S_{k}^{a\dagger}S_{k}^a\right. \nonumber \\
&+&\left. f_{k}^{s}\ a_kS_{k}^{s\dagger}+f_{k}^{s\ast}\ S_{k}^sa_k^{\dagger}\right\}.
\end{eqnarray}
The antisymmetric excitons are dark and decouple to the photons, while the
symmetric excitons are bright and coupled to the photons. Moreover the antisymmetric states, beside their localization, they
are metastable with damping rate of $\Gamma_a\approx 0$. The symmetric states form excitons and have finite life-time with damping
rate of $\Gamma_s\approx 2\Gamma_A$, two times the single excited atom one \cite{Ficek}.

At this point we replace the two-level operators by bosonic ones, that
is $S_{k}^{\nu}\rightarrow B_{k}^{\nu}$. In the next section we justify this
step and present a
bosonization procedure that goes beyond this simple replacement. In the strong coupling regime where the symmetric exciton and photon line widths are
smaller than the coupling parameter, the excitons and photons are
coherently mixed to form two polariton branches \cite{Kavokin}. The Hamiltonian is
diagonalized by using the upper and lower polariton operators
\begin{equation}
A_k^{\pm}=X_{k}^{\pm}\ B_{k}^s+Y_k^{\pm}\ a_k,
\end{equation}
which are a coherent superposition of symmetric excitons and photons. The mixing
amplitudes are defined by
\begin{equation}
X_{k}^{\pm}=\pm\sqrt{\frac{ D_k\mp\delta_k}{2 D_k}},\ \ \ Y_k^{\pm}=\frac{f_k^s}{\sqrt{2 D_k( D_k\mp\delta_k)}},
 \end{equation}
where $D_k=\sqrt{\delta_k^2+|f_k^s|^2}$, and the detuning is
$\delta_k=\frac{E_{ph}(k)-E_s(k)}{2}$. The diagonal Hamiltonian reads
\begin{equation}
H_{pol}=\sum_{k,r}E_{r}(k)\ A_k^{r\dagger}A_k^{r}+\sum_kE_{a}\
  B_{k}^{a\dagger}B_{k}^a,
\end{equation}
with the polariton dispersions
\begin{equation}
E_{\pm}(k)=\frac{E_{ph}(k)+E_s(k)}{2}\pm D_k.
\end{equation}
The exciton-photon detuning can be changed in order to control the mixing
amplitude around small wave number polaritons.

Now we present the results using some typical numbers. For the transition
energy we have $E_A=1.5\ eV$, for the lattice constant we use $a=1000\ \AA$,
and for the interatomic distance we have $R=100\ \AA$. The transition dipole
is $\mu=2.5\ e\AA$, the mode function is taken to be $u(b)=0.25$, and the
photon effective area is $S=\pi a^2$, with the effective dielectric constant
$\epsilon=2$.

In figure (3) we plot the polariton energies $E_{\pm}-E_A$, and the
excitation energies $E_s-E_A$ and $E_a-E_A$, as a function of the polarization angle
$\theta$ at zero wave number $k=0$. Here we take zero detuning between the
photon and the transition energy, that is $E_{ph}=E_A$. The symmetric states
have lower energy than the antisymmetric ones in the range from $\theta=0^o$ up to the magic
angle of $\theta_m=54.7^o$, and beyond $\theta_m$ up to $\theta=90^o$ the symmetric states
have higher energy. At $\theta_m$ we have $E_A=E_s=E_a$.

For later use we concentrate
in the dark states behavior. For polarization between $\theta=0^o$ and the
angle at which $E_a=E_+$, the dark states have energy higher than the upper
polariton branch, and they can intersect the upper polariton dispersion at a
finite wave number. For dark states between the angle at which $E_a=E_+$ and the
magic angle $\theta_m$ they fell in the energy gap between the
lower and upper polariton branches. Here the dark states have no intersection
with the polariton branches at any wave number. But dark states between the angle $\theta_m$ and
$\theta=90^o$ can intersect the lower polariton branch at a
finite wave number. The intersection point approaches small $k$ as $\theta$ approaches
$90^o$. This regime will be of interest later.

\begin{figure}[h!]
\centerline{\epsfxsize=8cm \epsfbox{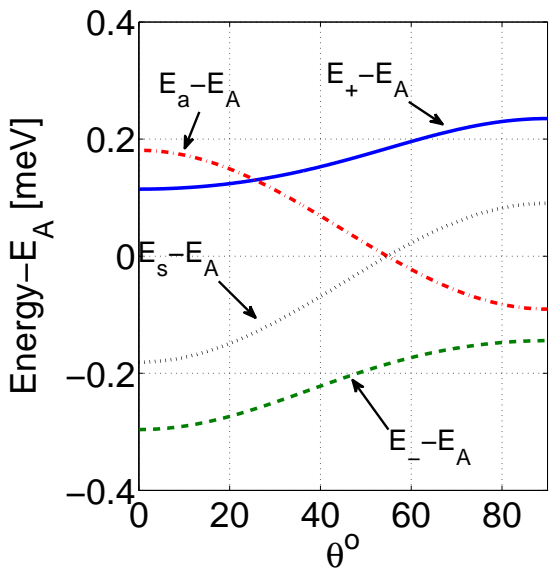}}
\caption{The energies $E_{\pm}-E_A$, $E_s-E_A$ and $E_a-E_A$ vs. $\theta$ at zero wave number $k=0$.}
\end{figure}

The polariton dispersions are plotted in figure (4) as a function of $k$. Here
we choose $\theta=80^o$, which is a reasonable angle in real experiments. We
plot also the photon dispersion and the symmetric-antisymmetric energies. The
dark state now intersects the lower polariton branch at a given wave number. In figure (5) we plot the excitation and photon fractions in the upper and lower polariton branches as a function of $k$, for $\theta=80^o$. In the upper branch, the full line is for
the photon fraction $|Y^+|^2$ and the dashed line for the excitation fraction
$|X^+|^2$, and vise versa in the lower branch.

\begin{figure}[h!]
\centerline{\epsfxsize=8cm \epsfbox{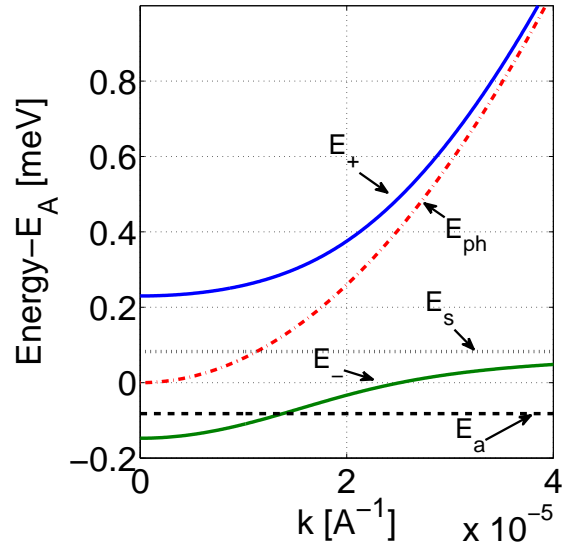}}
\caption{The energies $E_{\pm}-E_A$, $E_{ph}$, $E_s-E_A$ and $E_a-E_A$ vs. $k$ at the angle $\theta=80^o$.}
\end{figure}

\begin{figure}[h!]
\centerline{\epsfxsize=8cm \epsfbox{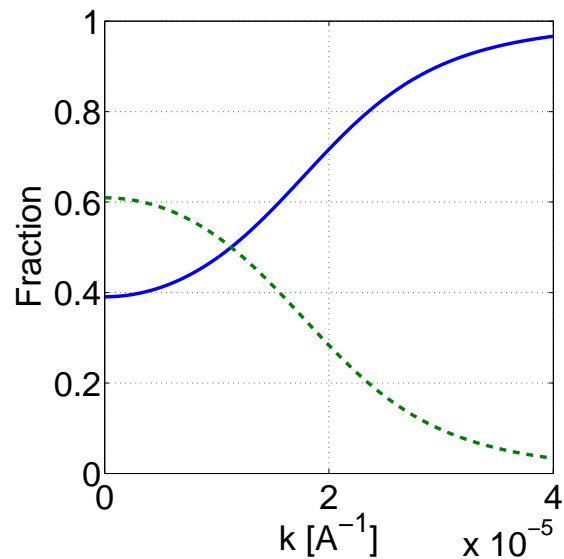}}
\caption{The
excitation and photon fractions in the upper and lower polariton branches
vs. $k$ at zero wave number $\theta=80^o$. In the upper branch, the full line is for
the photon fraction $|Y^+|^2$ and the dashed line for the excitation fraction
$|X^+|^2$, and vise versa in the lower branch.}
\end{figure}

\section{Kinematic Interactions}

Kinematic interactions are intensively studied by us in a previous work, and
found to be a significant nonlinear mechanism for collective excitations in
optical lattices \cite{Anderson,Toshich,Zoubi2005b}. In applying a bosonization
procedure we transform the excitation spin-half operators into boson ones with
additional terms that give rise to interactions between bosons \cite{Zoubi2014}. In solving the
scattering problem we derived an effective potential that we aim to
use in the following.

The operators $S_{n}^{\alpha\dagger}$ and $S_{n}^{\alpha}$ are of spin-half. They obey the fermi anti-commutation relation on the
same atom, that is
\begin{equation}
S_{n}^{\alpha}S_{n}^{\alpha\dagger}+S_{n}^{\alpha\dagger}S_{n}^{\alpha}=1\ ,\ S_{n}^{\alpha}S_{n}^{\alpha}=S_{n}^{\alpha\dagger}S_{n}^{\alpha\dagger}=0,
\end{equation}
and the bose commutation relation between different atoms, that is 
\begin{equation}
\left[S_{n}^{\alpha},S_{m}^{\beta\dagger}\right]=\left[S_{n}^{\alpha},S_{m}^{\beta}\right]=\left[S_{n}^{\alpha\dagger},S_{m}^{\beta\dagger}\right]=0,\
(n\neq m,\ \alpha\neq\beta).
\end{equation}
Then spin-half operators on a lattice have mixed statistics. They are
fermions on-site and bosons among different sites, where they usually termed
paulions.

We apply now the bosonization transformation \cite{Zoubi2014}
\begin{equation}
S_{n}^{\alpha}\rightarrow \left(1-B_{n}^{\alpha\dagger}B_{n}^{\alpha}\right)B_{n}^{\alpha}\ ,\ S_{n}^{\alpha\dagger}\rightarrow B_{n}^{\alpha\dagger}\left(1-B_{n}^{\alpha\dagger}B_{n}^{\alpha}\right),
\end{equation}
where $B_{n}^{\alpha\dagger}$ and $B_{n}^{\alpha}$ are boson operators with the commutation relation
$[B_{n}^{\alpha},B_{m}^{\beta\dagger}]=\delta_{nm}\delta_{\alpha\beta}$. This
transformation holds at low density of excitations, where the number of excitations is much smaller than that of lattice sites.

The bosonization yields interacting bosons with $H=H_0+H_I$, where the free boson
part reads
\begin{eqnarray}
H_0&=&\sum_k\left\{E_{ph}(k)\ a_k^{\dagger}a_k+E_{s}(k)\
  B_{k}^{s\dagger}B_{k}^s+E_{a}\
  B_{k}^{a\dagger}B_{k}^a\right. \nonumber \\
&+&\left. f_{k}^{s}\ a_kB_{k}^{s\dagger}+f_{k}^{s\ast}\ B_{k}^sa_k^{\dagger}\right\},
\end{eqnarray}
and the interaction part reads
\begin{equation} \label{HI}
H_I=U\sum_{n}\left\{B_{n}^{1\dagger}B_{n}^{1\dagger}B_{n}^1B_{n}^1+B_{n}^{2\dagger}B_{n}^{2\dagger}B_{n}^2B_{n}^2\right\},
\end{equation}
where $U$ is the kinematic effective potential.

In terms of
symmetric-antisymmetric operators we have
\begin{eqnarray}
H_I&=&\frac{U}{2}\sum_n\left\{B_{n}^{s\dagger}B_{n}^{s\dagger}B_{n}^sB_{n}^s+B_{n}^{a\dagger}B_{n}^{a\dagger}B_{n}^aB_{n}^a\right.
  \nonumber \\
&+&\left.B_{n}^{s\dagger}B_{n}^{s\dagger}B_{n}^aB_{n}^a+B_{n}^{a\dagger}B_{n}^{a\dagger}B_{n}^sB_{n}^s\right.
  \nonumber \\
&+&\left.4\ B_{n}^{s\dagger}B_{n}^{a\dagger}B_{n}^sB_{n}^a\right\}, \nonumber \\
\end{eqnarray}
and in momentum space we obtain
\begin{eqnarray}
H_I&=&\frac{U}{2N}\sum_{kk'\bar{k}}\left\{B_{k'-\bar{k}}^{s\dagger}B_{k+\bar{k}}^{s\dagger}B_{k'}^sB_{k}^s+B_{k'-\bar{k}}^{a\dagger}B_{k+\bar{k}}^{a\dagger}B_{k'}^aB_{k}^a\right.
  \nonumber \\
&+&\left.B_{k'-\bar{k}}^{s\dagger}B_{k+\bar{k}}^{s\dagger}B_{k'}^aB_{k}^a+B_{k'-\bar{k}}^{a\dagger}B_{k+\bar{k}}^{a\dagger}B_{k'}^sB_{k}^s\right.
  \nonumber \\
&+&\left.4\ B_{k'-\bar{k}}^{s\dagger}B_{k+\bar{k}}^{a\dagger}B_{k'}^sB_{k}^a\right\}.
\end{eqnarray}
We get kinematic interactions between symmetric states and between
antisymmetric states, as appear in the first two terms. Interestingly, the last
three terms represent interactions among symmetric and antisymmetric
states. Here two symmetric states can be scatter into two antisymmetric ones,
and vice versa. Moreover the last term represent scattering among
symmetric and antisymmetric states, and they will be our main concern in the
rest of the paper.

Writing the interaction Hamiltonian in terms of polariton operators yields
\begin{eqnarray}
H_I&=&\frac{U}{2N}\sum_{kk'\bar{k}}\left\{\sum_{\alpha\beta\gamma\delta}\left(X_{k'-\bar{k}}^{\delta}X_{k+\bar{k}}^{\gamma}X_{k'}^{\beta\ast}X_{k}^{\alpha\ast}\right)\right. \nonumber \\
&\times&\left.A_{k'-\bar{k}}^{\delta\dagger}A_{k+\bar{k}}^{\gamma\dagger}A_{k'}^{\beta}A_{k}^{\alpha}+B_{k'-\bar{k}}^{a\dagger}B_{k+\bar{k}}^{a\dagger}B_{k'}^aB_{k}^a\right.
  \nonumber \\
&+&\left.\sum_{\alpha\beta}\left[\left(X_{k'-\bar{k}}^{\beta}X_{k+\bar{k}}^{\alpha}\right)A_{k'-\bar{k}}^{\beta\dagger}A_{k+\bar{k}}^{\alpha\dagger}B_{k'}^aB_{k}^a\right.\right.
  \nonumber \\
&+&\left.\left.\left(X_{k'}^{\beta\ast}X_{k}^{\alpha\ast}\right)B_{k'-\bar{k}}^{a\dagger}B_{k+\bar{k}}^{a\dagger}A_{k'}^{\beta}A_{k}^{\alpha}\right.\right.
  \nonumber \\
&+&\left.\left.4\left(X_{k'-\bar{k}}^{\beta}X_{k'}^{\alpha\ast}\right)A_{k'-\bar{k}}^{\beta\dagger}B_{k+\bar{k}}^{a\dagger}A_{k'}^{\alpha}B_{k}^a\right]\right\}.
\end{eqnarray}
We concentrate only in the lower polariton branch, hence we can drop the
polariton branch indices. Furthermore, we interest in processes with small $k$, then one can neglect the dependence of
$X_k$ on $k$ and replace it by $X_0$ at $k=0$. We obtain
\begin{eqnarray}
H_I&=&\frac{\Delta}{2}\sum_{kk'\bar{k}}\left\{X_0^4\ A_{k'-\bar{k}}^{\dagger}A_{k+\bar{k}}^{\dagger}A_{k'}A_{k}+B_{k'-\bar{k}}^{\dagger}B_{k+\bar{k}}^{\dagger}B_{k'}B_{k}\right.
  \nonumber \\
&+&\left.X_0^2\
  A_{k'-\bar{k}}^{\dagger}A_{k+\bar{k}}^{\dagger}B_{k'}B_{k}+X_0^2\
  B_{k'-\bar{k}}^{\dagger}B_{k+\bar{k}}^{\dagger}A_{k'}A_{k}\right.
  \nonumber \\
&+&\left.4X_0^2\ A_{k'-\bar{k}}^{\dagger}B_{k+\bar{k}}^{\dagger}A_{k'}B_{k}\right\}.
\end{eqnarray}
Here, in order to simplify the notations, we used $A_k$ for the lower polariton
operator, and $B_k$ for the antisymmetric exciton operator. We also defined
$\Delta=\frac{U}{N}$, where the effective potential is given by
$U=\frac{4\pi\hbar^2}{ma^2}$, and in using $L=Na$, we get $\Delta=\frac{4\pi\hbar^2}{maL}$ \cite{Zoubi2014}.

\subsection{Dynamical Interactions}

Beside the kinematic interactions that result of quantum statistics, dynamical
interactions due to Coulomb forces are also of importance here. The dynamical
interactions have the form
\begin{equation}
H_{Dyn}=\sum_{nm}\sum_{\alpha\beta}V_{nm}^{\alpha\beta}\ S_n^{\alpha\dagger}S_m^{\beta\dagger}S_n^{\alpha}S_m^{\beta},
\end{equation}
where $\alpha\neq\beta$ for $n=m$. The interaction is among two excited atoms
and it is of higher order than the dipole-dipole interaction, e.g., of van der
Waals type. As the lattice constant is relatively large, we neglect dynamical
interactions among atoms at different unit cells, and we include only
interactions among two excited atoms on the same unit cell. Now we have
\begin{equation}
H_{Dyn}=V\sum_{n}\sum_{\alpha\beta}^{\prime}S_n^{\alpha\dagger}S_n^{\beta\dagger}S_n^{\alpha}S_n^{\beta},
\end{equation}
where the prime indicates that $\alpha\neq\beta$, with $V=V_{n}^{12}$, and we can
write $H_{Dyn}=2V\sum_{n}S_n^{1\dagger}S_n^{2\dagger}S_n^{1}S_n^{2}$. The state of two excited atoms in the same unit cell $n$ is defined by
$|e_n\rangle=|e_n^1,e_n^2\rangle$ with the energy $E_e=2E_A+E_B$, and the
binding energy $E_B=2V$.

This result is critical in the present work as it can exclude the possibility of double excitations
per site due to conservation of energy, where we have $(E_e\neq 2E_s, 2E_a, 2E_A)$. Hence in the next sections we can neglect the
scattering of two polaritons into the state $|e_n\rangle$ of two bound on-site excitations. But surely
they can be involved in other processes.

\section{Parametric excitation of the dark states}

It is a big experimental challenge to excite dark antisymmetric states as they are weakly coupled to the light. Usually they are populated through the decay of
higher states, or by multi-photon nonlinear processes. Here we exploit the result
of the previous section to suggest kinematic interactions as a mechanism for
significantly exciting such dark states. The above kinematic Hamiltonian
results in mixing interactions that
allow scattering among
bright polaritons and dark excitons.

We exploit the kinematic scattering of polaritons into dark excitons as a
mechanism for populating the dark states. In order to achieve this task in a
controllable way we present the following scenario that implies two external
sources in pump-probe experiment \cite{Boyd}. An external pump is used to strongly excite polaritons with
a fixed wave number, say $k$. A probe field is applied in order to
weakly excite dark states at a different wave number, say $k+q$ or $k-q$, or
both. This probe field is needed in order to stimulate scattering of bright
polaritons into these dark states. Therefore, for
the lower branch we consider only polaritons with a fixed wave
number $k$. The upper polariton branch is excluded due to conservation of
energy. As two polaritons of wave number $k$ can scatter into two dark
excitons, the conservation of momentum implies two dark excitons with momenta $k+q$
and $k-q$. Due to the fact that dark excitons are dispersion-less and on-site
localized, any pair of dark states is allowed in the scattering with wave numbers $k+q$ and $k-q$. Here we
consider the pump and probe fields to have the same energy $E$, where we choose
$E=E_{pol}(k)=E_a$. The probe field at wave numbers $k+q$ or $k-q$ are off
resonance with polaritons of energy $E_a$. On the other hand, the pump field can directly and
weakly excite dark excitons at $k$, and in the following we neglect the scattering of these
dark excitons into the $k+q$ and $k-q$ states. This configuration is used in order to
excite dark excitons with energies separated from the polariton in
an efficient and controllable manner, as
presented in figure (6).

\begin{figure}[h!]
\centerline{\epsfxsize=7cm \epsfbox{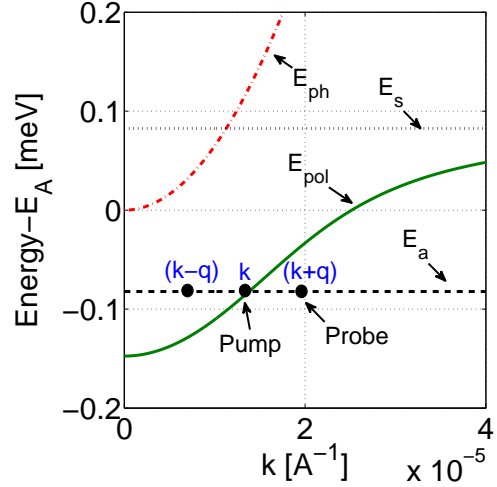}}
\caption{The energies $E_{pol}-E_A$, $E_{ph}$, $E_s-E_A$ and $E_a-E_A$ vs. $k$
  at the angle $\theta=80^o$. The cavity pump is at $k$ and the side probe at
  $k+q$. Two polaritons of wave number $k$ have stimulated scattering into two dark states
  at $k+q$ and $k-q$.}
\end{figure}

In the light of the above discussion, we consider the interaction
Hamiltonian with only $A_k$ polaritons, that is
\begin{eqnarray}\label{KinHam}
H_I&=&\frac{\Delta X_k^4}{2}\ A_{k}^{\dagger}A_{k}^{\dagger}A_{k}A_{k}+\frac{\Delta}{2}\sum_{kk'\bar{k}}B_{k'-\bar{k}}^{\dagger}B_{k+\bar{k}}^{\dagger}B_{k'}B_{k} \nonumber \\
&+&\frac{\Delta
  X_k^2}{2}\sum_{\bar{k}}\left\{A_{k}^{\dagger}A_{k}^{\dagger}B_{k+\bar{k}}B_{k-\bar{k}}+B_{k+\bar{k}}^{\dagger}B_{k-\bar{k}}^{\dagger}A_{k}A_{k}\right. \nonumber \\
&+&\left.4\
  A_{k}^{\dagger}B_{\bar{k}}^{\dagger}A_{k}B_{\bar{k}}\right\}.
\end{eqnarray}
The first term represent interactions among polaritons. The second term
gives interactions among dark excitons, and as dark states are weakly excited we
will neglect this term in the following. The second line includes scattering among polaritons and
dark excitons, where the first term describes scattering of two excitons into
two polaritons, and vice versa for the second one. The third line represents
the scattering of an exciton and a polariton. We concentrate in the interaction Hamiltonian 
\begin{eqnarray}
H_I&=&\frac{\Delta X_k^2}{2}\left\{X_k^2\
  A_{k}^{\dagger}A_{k}^{\dagger}A_{k}A_{k}+\sum_{p}\left[A_{k}^{\dagger}A_{k}^{\dagger}B_{k+p}B_{k-p}\right.\right. \nonumber \\
&+&\left.\left.B_{k+p}^{\dagger}B_{k-p}^{\dagger}A_{k}A_{k}+4\
  A_{k}^{\dagger}B_{p}^{\dagger}A_{k}B_{p}\right]\right\}.
\end{eqnarray}

We start by deriving equations of motion for the polariton and exciton operators
\begin{eqnarray}
i\hbar\frac{d}{dt}A_k&=&\left(E_{pol}(k)+\Delta X_k^4\
  A_{k}^{\dagger}A_{k}\right. \nonumber \\
&+&\left.2\Delta X_k^2\ \sum_{p}B_{p}^{\dagger}B_{p}\right)A_k \nonumber \\
&+&\Delta X_k^2\sum_{p}A_{k}^{\dagger}B_{k+p}B_{k-p}+F^{pump}_k, \nonumber \\
i\hbar\frac{d}{dt}B_{k+q}&=&\left(E_a+2\Delta X_k^2\
  A_{k}^{\dagger}A_{k}\right)B_{k+q} \nonumber \\
&+&\Delta X_k^2\ B_{k-q}^{\dagger}A_{k}A_{k}+F^{probe}_{k+q}, \nonumber \\
i\hbar\frac{d}{dt}B_{k-q}&=&\left(E_a+2\Delta X_k^2\
  A_{k}^{\dagger}A_{k}\right)B_{k-q} \nonumber \\
&+&\Delta X_k^2\ B_{k+q}^{\dagger}A_{k}A_{k}+F^{probe}_{k-q}.
\end{eqnarray}
We added three external fields, one field pumps the polaritons
through the waveguide at wave number $k$, and the probe fields excite weakly the dark
excitons at wave numbers $k+q$ and $k-q$. The pump and probe fields are taken to be of energy $E=\hbar\omega$, where
$F^{pump}_k=\tilde{F}^{pump}_ke^{-i\omega t}$, and
$F^{probe}_{p}=\tilde{F}^{probe}_{p}e^{-i\omega t}$. Hence we transfer into a system
that rotate at frequency $\omega$, by using $A_k=\tilde{A}_ke^{-i\omega t}$,
and $B_k=\tilde{B}_ke^{-i\omega t}$.

At this point we apply the mean field theory in taking the expectation value
of the equations, and by applying the factorization approximation
\begin{eqnarray}
\langle\tilde{A}_{k}^{\dagger}\tilde{A}_{k}\tilde{A}_k\rangle&=&\langle\tilde{A}_{k}^{\dagger}\tilde{A}_{k}\rangle\langle\tilde{A}_k\rangle, \nonumber \\
\langle\tilde{A}_{k}^{\dagger}\tilde{A}_{k}\tilde{B}_{k+q}\rangle&=&\langle\tilde{A}_{k}^{\dagger}\tilde{A}_{k}\rangle\langle\tilde{B}_{k+q}\rangle, \nonumber \\
\langle\tilde{A}_{k}^{\dagger}\tilde{A}_{k}\tilde{B}_{k-q}\rangle&=&\langle\tilde{A}_{k}^{\dagger}\tilde{A}_{k}\rangle\langle\tilde{B}_{k-q}\rangle, \nonumber \\
\langle\tilde{B}_{p}^{\dagger}\tilde{B}_{p}\tilde{A}_k\rangle&=&\langle\tilde{B}_{p}^{\dagger}\tilde{B}_{p}\rangle\langle\tilde{A}_k\rangle,
\nonumber \\
\langle\tilde{B}_{k+q}^{\dagger}\tilde{A}_{k}\tilde{A}_k\rangle&=&\langle\tilde{B}_{k+q}^{\dagger}\rangle\langle\tilde{A}_{k}\rangle^2, \nonumber \\
\langle\tilde{B}_{k-q}^{\dagger}\tilde{A}_{k}\tilde{A}_k\rangle&=&\langle\tilde{B}_{k-q}^{\dagger}\rangle\langle\tilde{A}_{k}\rangle^2,
\nonumber \\
\langle\tilde{A}_{k}^{\dagger}\tilde{B}_{k+p}\tilde{B}_{k-p}\rangle&=&\langle\tilde{A}_{k}^{\dagger}\rangle\langle\tilde{B}_{k+p}\rangle\langle\tilde{B}_{k-p}\rangle,
\end{eqnarray}
and we define $\langle\tilde{A}_k\rangle=\tilde{\cal
  A}_k$, $\langle\tilde{B}_{k}\rangle=\tilde{\cal B}_{k}$, $\langle\tilde{F}\rangle=\tilde{\cal
  F}$, $\langle\tilde{A}_{k}^{\dagger}\tilde{A}_{k}\rangle={\cal N}_k$, and
$\langle\tilde{B}_{k}^{\dagger}\tilde{B}_{k}\rangle={\cal I}_k$. For strong pump field we can neglect the contribution of the dark excitons
scattering into polaritons, hence we neglect terms that include dark
excitons in the first equation. Moreover, we define the renormalized exciton and
polariton energies by $\tilde{E}_{pol}(k)=E_{pol}(k)+\Delta X_k^4\ {\cal
  N}_k$, and $\tilde{E}_a=E_a+2\Delta X_k^2\ {\cal N}_k$. The mean field equations of motion read
\begin{eqnarray}
i\hbar\frac{d}{dt}\tilde{\cal A}_k&=&\left(\tilde{E}_{pol}(k)-E\right)\tilde{\cal A}_k+\tilde{\cal F}^{pump}_k, \nonumber \\
i\hbar\frac{d}{dt}\tilde{\cal B}_{k+q}&=&\left(\tilde{E}_a-E\right)\tilde{\cal B}_{k+q}+V\ \tilde{\cal
  B}_{k-q}+\tilde{\cal F}^{probe}_{k+q}, \nonumber \\
i\hbar\frac{d}{dt}\tilde{\cal B}_{k-q}&=&\left(\tilde{E}_a-E\right)\tilde{\cal
  B}_{k-q}+V\ \tilde{\cal B}_{k+q}+\tilde{F}^{probe}_{k-q},
\end{eqnarray}
where $V=\Delta X_k^2{\cal N}_{k}$.

At steady state, where $\frac{d}{dt}\tilde{\cal A}_k=\frac{d}{dt}\tilde{\cal
  B}_{k}=0$, we have
\begin{eqnarray}
\tilde{\cal B}_{k+q}&=&\frac{\left(E-\tilde{E}_a\right)\tilde{\cal
    F}^{probe}_{k+q}+V\ \tilde{\cal F}^{probe}_{k-q}}{\left(E-\tilde{E}_a\right)^2-V^2}, \nonumber \\
\tilde{\cal B}_{k-q}&=&\frac{\left(E-\tilde{E}_a\right)\tilde{\cal
    F}^{probe}_{k-q}+V\ \tilde{\cal F}^{probe}_{k+q}}{\left(E-\tilde{E}_a\right)^2-V^2}.
\end{eqnarray}
The fields have in general two resonances at $E_+=\tilde{E}_a+V$ and $E_-=\tilde{E}_a-V$. But for the specific case of $\tilde{\cal
  F}^{probe}_{k-q}=\tilde{\cal F}^{probe}_{k+q}$ we get only a single resonance at $E=\tilde{E}_a+V$.

We concentrate here in the case of one probe field at $k+q$, that is $\tilde{\cal F}^{probe}_{k-q}=0,\ \tilde{\cal
  F}^{probe}_{k+q}=\tilde{\cal F}^{probe}$, where we get
\begin{eqnarray}
\tilde{\cal B}_{k-q}&=&\frac{V\ \tilde{\cal F}^{probe}}{\left(E-\tilde{E}_a\right)^2-V^2}, \nonumber \\
\tilde{\cal B}_{k+q}&=&\frac{\left(E-\tilde{E}_a\right)\tilde{\cal
    F}^{probe}}{\left(E-\tilde{E}_a\right)^2-V^2},
\end{eqnarray}
with the intensities
\begin{eqnarray}
{\cal I}_{k+q}&=&\frac{\left(E-\tilde{E}_a\right)^2}{\left[\left(E-\tilde{E}_a\right)^2-\tilde{\Delta}^2{\cal N}_{k}^2\right]^2}|\tilde{\cal F}^{probe}_{k+q}|^2, \nonumber \\
{\cal I}_{k-q}&=&\frac{\tilde{\Delta}^2\ {\cal N}_{k}^{2}}{\left[\left(E-\tilde{E}_a\right)^2-\tilde{\Delta}^2{\cal N}_{k}^2\right]^2}|\tilde{\cal F}^{probe}_{k+q}|^2,
\end{eqnarray}
where $\tilde{\Delta}=\Delta X_k^2$. It is clear that we interestingly get two resonances at $E_+=\tilde{E}_a+V$ and
at $E_-=\tilde{E}_a-V$.

For the pumped polaritons we have
\begin{equation}
\tilde{\cal A}_{k}=\frac{\tilde{\cal F}^{pump}_{k}}{\left(E-\tilde{E}_{pol}(k)\right)},
\end{equation}
and
\begin{equation}
{\cal N}_{k}=\frac{|\tilde{\cal F}^{pump}_{k}|^2}{\left(E-\tilde{E}_{pol}(k)\right)^2}.
\end{equation}

In order to examine if such two resonances can be experimentally resolved we include the finite life time for both the exciton and the
polariton phenomenologically in changing the energies into complex parameters
$\tilde{E}_{pol}(k)\rightarrow \tilde{E}_{pol}(k)-i\hbar\Gamma_k^{pol}$, and
$\tilde{E}_a\rightarrow \tilde{E}_a-i\hbar\Gamma_a$, where $\Gamma_a$ is the
dark exciton damping rate, which is relatively a small
number. $\Gamma_k^{pol}$ is the polariton damping rate, that is defined by
$\Gamma_k^{pol}=\frac{1}{2}|X_k|^2\Gamma_s+\frac{1}{2}|Y_k|^2\Gamma_{ph}$, where $\Gamma_s$ is the symmetric exciton damping rate, and $\Gamma_{ph}$ is
the photon damping rate. The polariton and dark exciton average numbers are written as
\begin{equation}
{\cal N}_{k}=\frac{|\tilde{\cal
    F}^{pump}_{k}|^2}{\left(E-\tilde{E}_{pol}(k)\right)^2+\left(\hbar\Gamma_k^{pol}\right)^2},
\end{equation}
and
\begin{eqnarray}
{\cal I}_{k+q}&=&\frac{\left(E-\tilde{E}_a\right)^2+\left(\hbar\Gamma_a\right)^2}{\left(E-\tilde{E}_+\right)^2\left(E-\tilde{E}_-\right)^2}\ {\cal I}^{probe}, \nonumber \\
{\cal I}_{k-q}&=&\frac{\tilde{\Delta}^2\ {\cal
    N}_{k}^{2}}{\left(E-\tilde{E}_+\right)^2\left(E-\tilde{E}_-\right)^2}\ {\cal I}^{probe},
\end{eqnarray}
where ${\cal I}^{probe}=|\tilde{\cal F}^{probe}_{k+q}|^2$, and $\tilde{E}_{\pm}=\tilde{E}_a\pm\sqrt{\tilde{\Delta}^2{\cal N}_{k}^2-\hbar\Gamma_a^2}$.

We now illustrate the results in using the previous numbers. Moreover, we use for the photons $\hbar\Gamma_{ph}=10^{-10}\ eV$, for the
symmetric excitons $\hbar\Gamma_{s}=10^{-8}\ eV$, and for the antisymmetric
excitons $\hbar\Gamma_{a}=10^{-12}\ eV$. The lattice length is taken to be
$L=1\ cm$. We fix the polarization angle to $\theta=80^o$. We pump lower
branch polaritons at energy $E_{pol}(k)=E_a$, which is at wave number $k\approx
1.4\times 10^{-5}\ \AA^{-1}$. To calculate the effective interaction parameter
we need to evaluate the polariton effective mass, which is of the order of the cavity
photon effective mass, hence we have $mc^2\approx \hbar c q_0\sqrt{\epsilon}$,
and for $E_A=E_{ph}(0)$ we obtain $q_0=\sqrt{\epsilon} E_A/(\hbar c)$. Here we
have $q_0\approx 10^{-3}\ \AA^{-1}$, $mc^2\approx 3\ eV$, and for the
interaction parameter we get $\Delta\approx 1.6\times 10^{-4}\ eV$. At this $k$ we get $|X_k|^2\approx 0.56$ hence we have $\tilde{\Delta}\approx
9.1\times 10^{-5}\ eV$. The pumped polaritons are assumed to be with average
number of ${\cal N}_{k}\sim 1$. In figure (7) we plot the scaled intensity
${\cal I}_{k-q}/{\cal I}^{probe}$ as a function of $E-E_a$. We pump at $k$ and probe at
$k+q$ of energy $E_a$, and hence the dark excitons get excited at $k-q$ with blue
shift energy. As dark excitons have small $\Gamma_a$, that is much smaller than
the interaction parameter $\tilde{\Delta}$ the spectra split into two
peaks at $\tilde{E}_+$ and $\tilde{E}_-$. We get parametric amplification of
the dark excitons both at $k+q$ and $k-q$ with blue shifts, which can be
experimentally resolved.

\begin{figure}[h!]
\centerline{\epsfxsize=8cm \epsfbox{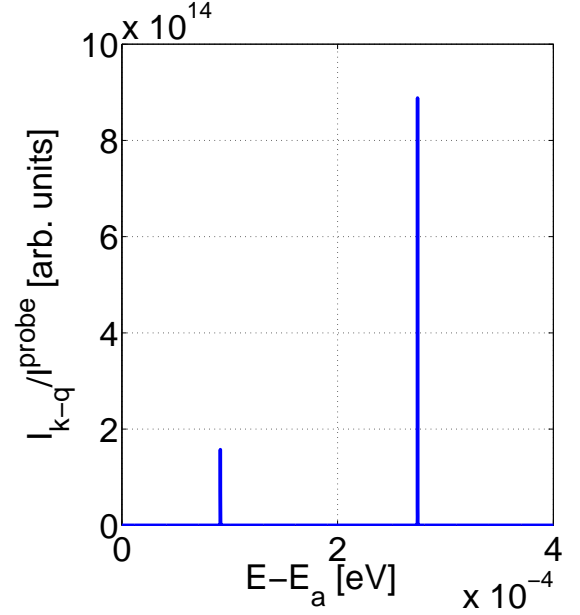}}
\caption{The scaled intensity
${\cal I}_{k-q}/{\cal I}^{probe}$ vs. $E-E_a$. The peaks are easily resolved.}
\end{figure}

\section{Dark Bogolon-Excitons}

We now aim to get a much deep understanding of the previous section results
for the
dark states. In order to achieve this target we follow a different direction for
treating pair excitations of dark excitons in adopting Bogoliubov mean
field theory \cite{Fetter}.

We start again from the Hamiltonian (\ref{KinHam}). Now we assume the strong pump of the polaritons at wave number $k$ to be a
classical field, which fits with the previous mean field theory
treatment. Hence the first term becomes a number. As before, the dark states are
weakly excited then we can neglect the second term of their mutual
interactions. These assumptions are in the spirit of Bogoliubov mean field
theory \cite{Fetter}. In using
$\langle A_{k}\rangle=\sqrt{\cal N}e^{-i\omega t}$, the total Hamiltonian reads
\begin{eqnarray}
H&=&\sum_{p}\left\{\frac{1}{2}\left(E_a+2\Delta{\cal N}
    X_k^2\right)\left(B_{k+p}^{\dagger}B_{k+p}+B_{k-p}^{\dagger}B_{k-p}\right)\right.
  \nonumber \\
&+&\left.\frac{\Delta{\cal N}
  X_k^2}{2}\left(e^{i2\omega t}\ B_{k+p}B_{k-p}+e^{-i2\omega t}\ B_{k+p}^{\dagger}B_{k-p}^{\dagger}\right)\right\}. \nonumber \\
\end{eqnarray}
We transform into rotating frame in applying the transformation $U=e^{-iSt/\hbar}$,
hence we have $H\rightarrow U^{\dagger}HU-S$, with
$S=E\left(B_{k+p}^{\dagger}B_{k+p}+B_{k-p}^{\dagger}B_{k-p}\right)$, to get
\begin{eqnarray}
H&=&\frac{1}{2}\sum_{p}\left\{\left(\tilde{E}_a-E\right)\left(\tilde{B}_{k+p}^{\dagger}\tilde{B}_{k+p}+\tilde{B}_{k-p}^{\dagger}\tilde{B}_{k-p}\right)\right.
  \nonumber \\
&+&\left.V\left(\tilde{B}_{k+p}\tilde{B}_{k-p}+\tilde{B}_{k+p}^{\dagger}\tilde{B}_{k-p}^{\dagger}\right)\right\},
\end{eqnarray}
where $\tilde{E}_a=E_a+2V$, and $V=\Delta{\cal N}X_k^2$.

The form of the Hamiltonian appeals to the Bogoliubov transformation \cite{Fetter}
\begin{eqnarray}
\alpha_p&=&u\ \tilde{B}_{k+p}+v\ \tilde{B}_{k-p}^{\dagger},  \nonumber \\
\beta_p&=&u\ \tilde{B}_{k-p}+v\ \tilde{B}_{k+p}^{\dagger}, 
\end{eqnarray}
with
\begin{equation}
\left[\alpha_p,\alpha_p^{\dagger}\right]=\left[\beta_p,\beta_p^{\dagger}\right]=1,
\end{equation}
and
\begin{equation}
\left[\alpha_p,\beta_p^{\dagger}\right]=\left[\beta_p,\alpha_p^{\dagger}\right]=\left[\alpha_p,\beta_p\right]=\left[\alpha_p^{\dagger},\beta_p^{\dagger}\right]=0,
\end{equation}
where $u^2-v^2=1$. The inverse transformation is
\begin{eqnarray}
\tilde{B}_{k+p}&=&u\ \alpha_p-v\ \beta_p^{\dagger},  \nonumber \\
\tilde{B}_{k-p}&=&u\ \beta_p-v\ \alpha_p^{\dagger}.
\end{eqnarray}
In terms of the new operators the Hamiltonian is reads
\begin{eqnarray}
H&=&\sum_p\left\{\left(\tilde{E}_a-E\right)v^2-V\
  uv\right. \nonumber \\
&+&\left.\left[\frac{1}{2}\left(\tilde{E}_a-E\right)\left(u^2+v^2\right)-V\ uv\right]\left(\alpha_p^{\dagger}\alpha_p+\beta_p^{\dagger}\beta_p\right)\right.
  \nonumber \\
&+&\left.\left[\frac{V}{2}\ \left(u^2+v^2\right)-\left(\tilde{E}_a-E\right)uv\right]\left(\alpha_p\beta_p+\alpha_p^{\dagger}\beta_p^{\dagger}\right)\right\}. \nonumber \\
\end{eqnarray}
We need the second line to vanish, then we can choose $u$ and $v$ such that
\begin{equation}
\frac{V}{2}\ \left(u^2+v^2\right)=\left(\tilde{E}_a-E\right)uv.
\end{equation}
We take $u$ and $v$ to be real, and we choose $u>0$, hence from $u^2-v^2=1$,
we use $u=\cosh t$ and $v=\sinh t$, and the above relation gives $\tanh
2t=V/\left(\tilde{E}_a-E\right)$, and we get the solution
\begin{equation}
u^2=\frac{1}{2}\left(\frac{\tilde{E}_a-E}{2\tilde{E}_0}+1\right),\ v^2=\frac{1}{2}\left(\frac{\tilde{E}_a-E}{2\tilde{E}_0}-1\right),
\end{equation}
where
\begin{equation}
\tilde{E}_0=\frac{1}{2}\sqrt{\left(\tilde{E}_a-E\right)^2-V^2}.
\end{equation}
Using this solution, the Hamiltonian is given by
\begin{equation}
H=\sum_p\left\{\tilde{E}_0-\frac{\tilde{E}_a-E}{2}+\tilde{E}_0\left(\alpha_p^{\dagger}\alpha_p+\beta_p^{\dagger}\beta_p\right)\right\}.
\end{equation}

As $\alpha_p=\beta_{-p}$ and $\beta_{p}=\alpha_{-p}$, we define the operators
\begin{eqnarray}\label{CTrans}
C_p&=&u\ \tilde{B}_{k+p}+v\ \tilde{B}_{k-p}^{\dagger},  \nonumber \\
C_{-p}&=&u\ \tilde{B}_{k-p}+v\ \tilde{B}_{k+p}^{\dagger},
\end{eqnarray}
then we can write
\begin{equation}
H=\sum_p\left\{\frac{1}{2}\left(\bar{E}_0-\tilde{E}_a+E\right)+\bar{E}_0\ C_p^{\dagger}C_p\right\},
\end{equation}
where here $\bar{E}_0=2\tilde{E}_0$. The diagonal Hamiltonian represents dark state bogolons, in which two dark states of momenta $k+p$ and
$k-p$ are strongly correlated. These elementary excitations are termed here bogolon-excitons and found to be dispersion-less with energy $\bar{E}_0$.

We use a probe field that weakly excites the dark excitons at $k+q$, then
the Bogoliubov mean field equations of motion in rotating frame for
the new operators are
\begin{eqnarray}
i\hbar\frac{d}{dt}{\cal C}_{k+q}&=&\bar{E}_0\ {\cal C}_{k+q}+u\ \tilde{\cal F},  \nonumber \\
i\hbar\frac{d}{dt}{\cal C}_{k-q}&=&\bar{E}_0\ {\cal C}_{k-q}-v\ \tilde{\cal F}.
\end{eqnarray}
At steady state
\begin{equation}
{\cal C}_{k+q}=-\frac{u\ \tilde{\cal F}}{\bar{E}_0},\ {\cal C}_{k-q}=\frac{v\ \tilde{\cal F}}{\bar{E}_0}.
\end{equation}
The results exactly confirm the previous section ones, which can be obtained
in using the inverse of the (\ref{CTrans}) transformation.

We conclude that in the case of a single probe field at $k+q$,
the two dark excitons, which appear through the scattering of two $k$
polaritons, are spontaneously correlated and form two coherent states that are
separated by the
interaction energy $V$ in the optical spectra.

\section{Conclusions}

In the present paper we showed how coherent states of
electronic excitations are spontaneously formed only through resonant dipole-dipole
interactions and in exploiting the lattice symmetry, without the aid of additional
external fields. The study is presented for an optical super-lattice with two
atoms per unit cell in which dark and bright excitons appear. In localizing
the one dimensional atomic super-lattice parallel to a nanophotonic waveguide
the bright excitons are strongly coupled to the waveguide photons to form
polaritons as the system natural collective excitations.

Excitons behave as bosons at low intensity of excitations, but as the atomic transitions are of two-level systems a
mechanism is required to forbid two excitations from being localized on the
same atom. A bosonization procedure from two-level systems into bosons was
suggested by us in other work that excludes double excitation at the same atom
by including kinematic interactions among excitons. In adopting this procedure
for the present atomic super-lattice the kinematic interactions interestingly
couple dark and bright excitons.

We used the kinematic interactions in order to excite dark excitons in an
efficient and controllable way. We suggested a pump-probe experiment in which two
pumped polaritons of a given momentum $k$ can be scattered into two dark
excitons of momenta $k+p$ and $k-p$, where the process
is induced by a probe field. The two dark excitons are found to be strongly
correlated and can be represented as dark bogolon-excitons.

Bogolon-excitons can be implemented as qubits for quantum information
processing, and their long life-time allow them to serve as memory states. The
above parametric amplification scenario can be used to write into these states,
and the addition of asymmetry into the system can be used to read
out. Furthermore, achieving a fixed population of dark bogolon-excitons in a controllable way is an important step toward
achieving the long awaited BEC of excitons.

The consideration in the present paper event hough presented for atomic
super-lattice can be also realized for a one
dimensional organic crystal with two molecules per unit cell. Here one get
an anisotropic crystal and the orientation
of the molecules relative to the axis gives rise to two Davydov exciton
branches, in which both can be bright or one bright and the other dark. This
fact can enrich the present physical properties as the polarization direction provides
an extra degree of freedom.

\end{document}